\documentclass[trackchanges,twocolumn]{aastex701}

\usepackage{amsmath}
\usepackage{bm}
\usepackage{booktabs}
\usepackage{placeins}
\usepackage{hyperref} 
\newcommand{\Msun}{M_{\odot}}
\newcommand{\Req}{R_{\rm eq}}

\newcommand{\ee}{\mathrm{e}}

\shorttitle{Analytic OS Profiles}
\shortauthors{Yang and Zhao}

\begin{document}

\author[0009-0009-1295-6556]{Xi Yang}
\affiliation{School of Physics, Peking University, Beijing 100871, China}\email{2500011585@stu.pku.edu.cn}

\author[0009-0003-6348-7143]{Tong Zhao}
\affiliation{School of Physics, Peking University, Beijing 100871, China}\email{tongzhao2614@pku.edu.cn}

\author[0000-0002-9042-3044]{Renxin Xu}
\affiliation{School of Physics, Peking University, Beijing 100871, China}\email{r.x.xu@pku.edu.cn}

\title{A Refined Analytic Oblate--Schwarzschild Model for Thermal X-Ray Pulse Profiles of Rotating Neutron Stars}

\begin{abstract}
Thermal X-ray pulse profiles constrain neutron-star masses and radii through the effects of gravitational light bending and rotation on surface emission. The precision anticipated from the enhanced X-ray Timing and Polarimetry (eXTP) mission motivates more accurate analytic light-bending formulae that retain the efficiency needed for repeated model evaluations. We derive a refined analytic flux expression within the oblate--Schwarzschild (OS) approximation that combines bending, propagation-delay, and lensing approximations with a spheroidal surface. We compare its profiles with numerical Schwarzschild and second-order Hartle--Thorne ray tracing on the same spheroidal surfaces. The comparisons use a selected two-spot configuration at 300 Hz with isotropic blackbody emission and three choices for surface eccentricity: a freely specified value, the Maclaurin relation for a uniformly rotating Newtonian fluid, and a fit to rotating-neutron-star surfaces. Across these cases, the refined OS profiles have RMS flux differences of $1.76-1.98\%$ relative to the Hartle--Thorne reference, compared with $6.50-21.39\%$ for the spherical Schwarzschild-plus-Doppler (S+D) model. All differences are normalized to the phase-averaged reference flux. The corresponding differences between Schwarzschild and Hartle--Thorne ray tracing remain below $1.1\%$. These comparisons separate errors in analytic photon transfer from the effects of surface deformation and the rotating exterior spacetime. For the configurations examined, accounting for oblateness substantially improves agreement with ray tracing. The model retains the speed of an analytic calculation while making the physical origin of profile changes explicit.
\end{abstract}
\section{Introduction}
Gravitational light bending connects radiation emitted near a compact object to the images, spectra, and light curves measured by a distant observer. In Schwarzschild spacetime, the light-bending equation relates the local emission angle $\alpha$, measured from the radial direction, to the angle $\psi$ between the emission radius vector and the observer's direction. This relation determines which emission directions reach the observer, while its angular derivative enters the lensing factor that converts an emitting area into an observed solid angle~\citep{Pechenick1983,Beloborodov2002,Poutanen2020}. Accurate evaluation of both quantities is therefore essential for predicting the observed flux from neutron-star surfaces and accretion disks around black holes.

For neutron-star surfaces, this mapping is central to thermal X-ray pulse-profile modeling. Thermal radiation probes the surface temperature, atmospheric properties, and internal structure of a neutron star~\citep{Helfand1980,Zavlin2009}. Localized hot regions can arise from heating of the magnetic polar caps by particles returning from the magnetosphere~\citep{HardingMuslimov2001}, while a nonuniform surface temperature also produces rotational modulation in cooling stars~\citep{Zavlin2009}. Light bending makes part of the far side of a neutron star visible and changes the pulse amplitude and shape. For broad, nearly isotropic emission, it generally reduces the modulation~\citep{Pechenick1983,Beloborodov2002}. The resulting profiles depend on compactness, viewing geometry, and the angular distribution of the emitted intensity. Doppler shifts, aberration, and propagation delays provide additional sensitivity to the stellar radius and rotation~\citep{PoutanenBeloborodov2006}.  Combining profiles at different photon energies helps constrain these correlated parameters~\citep{PsaltisEtAl2014}. Analyses of thermal pulsations from pulsars with the \textit{Neutron Star Interior Composition Explorer} (NICER) have demonstrated the use of this approach to infer neutron-star masses and radii and constrain the dense-matter equation of state~\citep{Miller2019,Riley2019}.

Pulse profiles can also constrain stellar properties when the emitting regions are produced by thermonuclear burning or accretion. In models of X-ray burst oscillations, rotating brightness asymmetries generate oscillations near the stellar spin frequency. As in thermal polar-cap models, gravitational bending controls the visibility and apparent area of the emitting region, so the waveform carries information about compactness together with the spot and observer inclinations.~\citet{Lo2013} used simulated energy-resolved burst waveforms to study the mass and radius constraints obtainable from this information, including their dependence on geometry, photon counts, and background.~\citet{MillerLamb2015} extended such inference calculations to oblate stars, showing the importance of including rotational deformation in burst waveform models. Light bending also enters models of persistent pulsations from accreting millisecond pulsars, where magnetically channeled accretion creates bright surface regions~\citep{PoutanenBeloborodov2006}.

The transfer of radiation from accretion disks around black holes involves a similar mapping between emission and observation. For a disk around a Schwarzschild black hole, bent photon paths determine the apparent disk image, including the distorted view of its far side and secondary images~\citep{Luminet1979}. Together with gravitational redshift and the Doppler effects of orbital motion, lensing determines the relative contributions of different disk regions to the observed continuum and line spectra~\citep{Cunningham1975,Poutanen2020}.~\citet{Poutanen2020} explicitly demonstrated the use of an analytic Schwarzschild bending approximation to calculate disk line profiles. For rotating black holes, the corresponding transfer calculation requires photon trajectories in Kerr spacetime, as in~\citet{Cunningham1975}. Both surface-emission and disk-emission models thus rely on accurate photon transfer to connect source properties to observables.

Using these models for parameter inference requires repeated ray tracing over extended emitting regions and photon energies, as well as rotational phase for pulse profiles. Even in Schwarzschild spacetime, the exact light bending must be evaluated by solving geodesic equations, or represented by interpolation tables. Analytic approximations reduce this computational cost and make parameter dependencies explicit.~\citet{Beloborodov2002} introduced a simple approximate relation between $\cos\alpha$ and $\cos\psi$, which underlies useful analytic descriptions of pulsar profiles~\citep{PoutanenBeloborodov2006}.~\citet{Poutanen2020} subsequently improved the accuracy of the bending relation and its associated lensing factor over a wider range of emission radii and observer angles. Because the observed flux depends on the angular derivative as well as the bending angle, the accuracy of both quantities must be assessed when using these approximations in waveform calculations.

The need to control these approximation errors becomes more pressing as observational precision improves. The higher photon statistics anticipated from the enhanced X-ray Timing and Polarimetry (eXTP) mission are expected to improve the precision of neutron-star mass and radius measurements through pulse-profile modeling~\citep{eXTP2025}. To exploit this improvement, waveform models must keep errors from approximate photon propagation small enough that they do not limit the precision of the inferred parameters. As statistical uncertainties decrease, residual errors in the predicted pulse shape can introduce significant biases. Analytic models therefore need accurate light-bending and lensing formulae whose predicted fluxes can be checked against numerical ray tracing, while remaining fast enough for repeated parameter inference.

Meeting these accuracy requirements also calls for a consistent treatment of the emitting surface of a rotating neutron star. The Schwarzschild-plus-Doppler (S+D) approximation includes special-relativistic surface motion but assumes a spherical star. Rotational oblateness changes the local emission radius, proper area, surface normal, and visibility of each emitting element. Ray-tracing studies show that these geometric effects can exceed the corrections associated with the rotating exterior spacetime~\citep{Cadeau2007,PsaltisOzel2014}. The oblate-Schwarzschild (OS) approximation captures much of this effect by placing an oblate surface in a Schwarzschild exterior~\citep{Morsink2007}. Combining this geometry with accurate analytic bending and lensing formulae provides a way to improve both surface and light-bending treatments within an efficient
pulse-profile model.

In this work, we derive a refined analytic OS model for thermal X-ray pulse profiles. We combine a spheroidal surface and its local normal with the bending and lensing approximations of~\citet{Poutanen2020}, special-relativistic effects, finite emitting regions, and propagation delays. The observed flux is expressed in terms of redshift, Doppler, lensing, projected-area, and emission factors. We compare the model with numerical Schwarzschild ray tracing on the same surface and with Hartle--Thorne ray tracing through second order in rotation~\citep{HartleThorne1968}. Surface eccentricity is specified directly, calculated from the Maclaurin relation for a uniformly rotating Newtonian fluid, or obtained from the neutron-star surface fit of~\citet{Morsink2007}. Empirical universal relations connect the stellar inputs to the angular momentum and quadrupole used in the Hartle--Thorne calculation~\citep{BreuRezzolla2016,YagiYunes2013}. Comparing these profiles separates the effects of surface deformation, approximate photon transfer, and the rotating exterior spacetime.

Section~\ref{sec:model} presents the flux calculation, oblate geometry,
photon transfer, and stellar parameter prescriptions.
Section~\ref{sec:results} compares the models for a two-spot configuration
at 300 Hz under the three eccentricity prescriptions.
Section~\ref{sec:conclusions} summarizes the conclusions.

\section{Analytic Model} \label{sec:model}

We use a common observed-flux expression to separate changes in surface
geometry from changes in photon propagation. We first give the spherical
S+D flux, then introduce the oblate surface, the analytic transfer terms,
and the choices of surface shape and spacetime parameters.

\subsection{Spherical S+D baseline and observed flux}\label{subsec:flux}

The Schwarzschild-plus-Doppler (S+D) model provides the spherical baseline
against which we isolate surface-deformation effects. It places a
surface of radius $\Req$ in a Schwarzschild exterior and includes the
special-relativistic motion of the hot spot. Let $R$ be the local
emission radius,
$u=2GM/(c^2R)$ its compactness, $\theta$ the colatitude, $\phi$ the rotational phase, and $\Omega$ the
stellar angular frequency.  The surface speed measured by a local static
observer and the Doppler factor are
\begin{equation}
      \beta_{\rm rot}=\frac{\Omega R\sin\theta}{c\sqrt{1-u}},\qquad
      \delta=\frac{1}{\Gamma[1-\beta_{\rm rot}\cos\xi]},
      \label{eq:doppler}
\end{equation}
where $\Gamma=(1-\beta_{\rm rot}^2)^{-1/2}$ and $\xi$ is the angle between
the photon direction and the surface velocity.  If $\beta$ is the emission
angle relative to the local surface normal, aberration gives
$\cos\beta'=\delta\cos\beta$ in the comoving frame, while
$E=\delta\sqrt{1-u}\,E'$ relates observed and emitted photon energies.

For comoving specific intensity $I'_{E'}$ and distance $D$, the observed
spectral flux from a surface element is
\begin{equation}
      dF_E=\sqrt{1-u}\,\delta^4 I'_{E'}(\cos\beta')
      \mathcal{J}\cos\beta\frac{dS}{D^2},
      \label{eq:masterflux}
\end{equation}
where $dS$ is the proper surface element and $\mathcal{J}$ is the
light-bending Jacobian.  In the spherical S+D limit, $R=\Req$,
$\beta=\alpha$, with $\alpha$ measured from the radial direction, and
$dS=\Req^2\sin\theta\,d\theta\,d\phi$.  Our OS calculation retains
Equation~(\ref{eq:masterflux}) but evaluates its factors on an oblate
surface.  The present comparisons use isotropic blackbody emission.

\subsection{Oblate surface geometry}\label{subsec:geometry}

\begin{figure*}[t]
    \centering
    \includegraphics[width=0.73\textwidth]{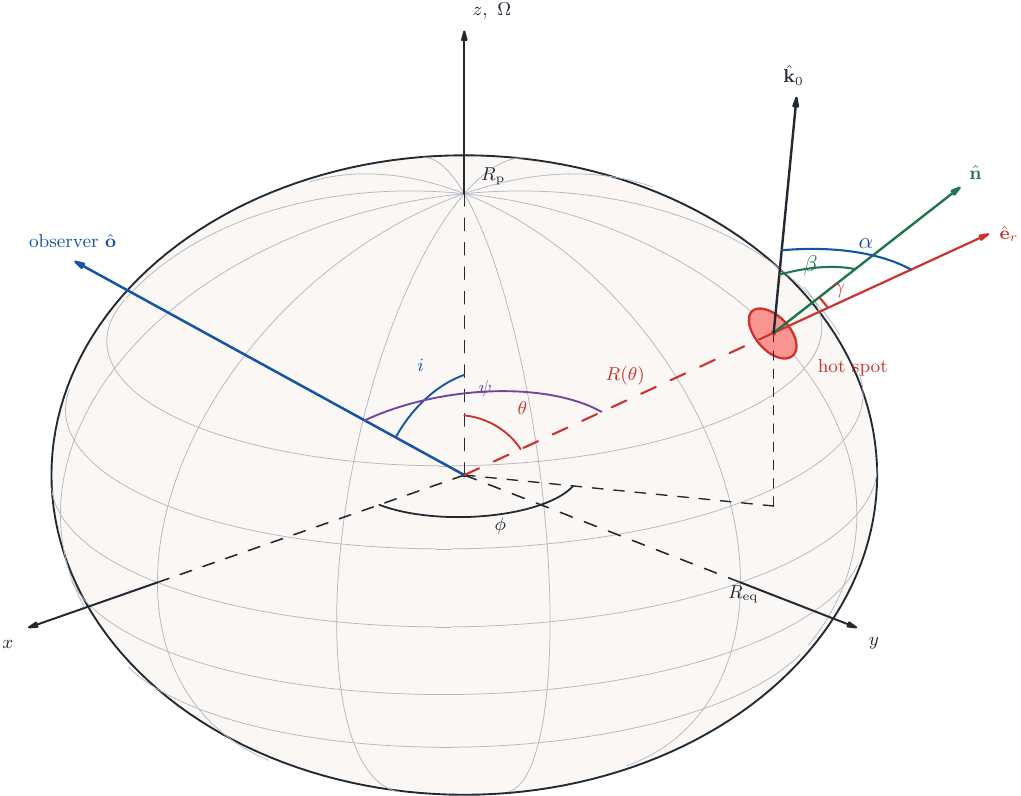}
    \caption{Oblate-star geometry.  The spot at $(\theta,\phi)$ is seen
            at inclination $i$, with $\psi$ between its radial direction
            and the observer direction $\hat{\mathbf o}$.  At the spot,
            $\alpha$ and $\beta$ are the angles between
            $(\hat{\mathbf k}_0,\hat{\mathbf e}_r)$ and
            $(\hat{\mathbf k}_0,\hat{\mathbf n})$, respectively;
            $\gamma$ is the signed meridional tilt of
            $\hat{\mathbf n}$ from $\hat{\mathbf e}_r$ (shown positive).
            The drawing is schematic.}
      \label{fig:stellargeometry}
\end{figure*}

We represent the stellar surface by an oblate spheroid. In Newtonian gravity,
the Maclaurin spheroid is an exact equilibrium solution for a uniformly
rotating, incompressible fluid. For a separate illustrative 300 Hz neutron-star model,
the best-fit spheroid reproduces the radial shape prescribed by
\citet{Morsink2007} to fractional accuracy of order $10^{-3}$
(Section~\ref{subsec:shape}). The shape enters the flux through its
latitude-dependent emission radius, proper area, and surface normal.
Figure~\ref{fig:stellargeometry} illustrates the local geometry.
For the OS model, $\Req$ is the equatorial circumferential radius and $i$ is
the observer inclination.  The angle $\psi$ between the radial direction
and the observer at rotational phase $\phi$ is
\begin{equation}
 \cos\psi=\cos i\cos\theta+\sin i\sin\theta\cos\phi.
 \label{eq:psi}
\end{equation}
The exact spheroidal surface used in the calculation is
\begin{equation}
 R(\theta)=\frac{\Req\sqrt{1-e_\star^2}}
 {\sqrt{1-e_\star^2\sin^2\theta}},
 \label{eq:spheroid}
\end{equation}
where $e_\star=[1-(R_{\rm p}/\Req)^2]^{1/2}$ is the geometric eccentricity.
The compactness $u$ in Equation~(\ref{eq:masterflux}) therefore varies with
colatitude through $R(\theta)$.

The meridional slope determines the signed tilt
$\gamma=\arctan f(\theta)$ of the outward surface normal from the
radial direction; positive $\gamma$ points toward decreasing colatitude:
\begin{equation}
 f(\theta)=\frac{1}{\sqrt{1-u(\theta)}}
 \frac{1}{R(\theta)}\frac{dR}{d\theta}, \qquad
 \cos\gamma=\frac{1}{\sqrt{1+f^2}}.
 \label{eq:slope}
\end{equation}
The proper surface element is
\begin{equation}
 dS=R^2\sin\theta\sqrt{1+f^2}\,d\theta\,d\phi.
 \label{eq:area}
\end{equation}
Let $\delta_{\rm m}$ be the angle between the tangential projection
of the initial photon direction onto the plane perpendicular to
$\hat{\mathbf e}_r$ and $-\hat{\mathbf e}_{\theta}$, the direction of
decreasing colatitude.  If $\alpha$ is measured from the radial
direction, the photon angle $\beta$ from the true surface normal is
\begin{equation}
 \cos\beta=\cos\alpha\cos\gamma
 +\sin\alpha\sin\gamma\cos\delta_{\rm m}.
 \label{eq:beta}
\end{equation}
The visibility condition is $\cos\beta'>0$ in the comoving frame, rather
than the spherical condition $\cos\alpha>0$.

\subsection{Light bending and the lensing Jacobian}\label{subsec:bending}

The analytic photon transfer uses approximations to Schwarzschild light
bending and its lensing Jacobian in place of tracing each ray numerically.
For an initially outgoing photon with impact parameter $b$, the exact
bending relation is
\begin{equation}
 \psi(b,R)=b\int_R^\infty\frac{dr}{r^2}
 \left[1-\frac{b^2}{r^2}\left(1-\frac{2GM}{rc^2}\right)\right]^{-1/2},
 \label{eq:exactbend}
\end{equation}
with
\begin{equation}
 \sin\alpha=\frac{b}{R}\sqrt{1-u}.
\end{equation}
Evaluating Equation~(\ref{eq:exactbend}) at every surface point would require
numerical inversion. Defining $x=1-\cos\alpha$ and $y=1-\cos\psi$, we instead use
\begin{equation}
 x=(1-u)y\left[1+\frac{u^2y^2}{112}
 -\frac{\ee\,uy}{100}\left(\ln(1-y/2)+y/2\right)\right],
 \label{eq:poutanenbend}
\end{equation}
where $\ee$ is Euler's number and must not be confused with $e_\star$.
The corresponding lensing Jacobian is
\begin{align}
 \mathcal{J}&=\left|\frac{\partial\cos\alpha}
 {\partial\cos\psi}\right|_R=(1-u)\mathcal{D},\\
 \mathcal{D}&=1+\frac{3u^2y^2}{112}
 -\frac{\ee\,uy}{100}\left[2\ln(1-y/2)
 +\frac{y(1-3y/4)}{1-y/2}\right].
 \label{eq:jacobian}
\end{align}
Equations~(\ref{eq:poutanenbend}) and (\ref{eq:jacobian}) follow the
high-accuracy approximation of \citet{Poutanen2020}.  Initially ingoing rays
near the limb are handled with a separate local expansion, summarized in
Appendix~\ref{app:ingoing}.

\subsection{Propagation delay, finite spots, and phase}\label{subsec:delay}

Photon travel-time shifts affect the observed phase, while integration
over finite spots accounts for emission across each cap.
For outgoing rays we use the analytic delay approximation
\citep{PoutanenBeloborodov2006}
\begin{equation}
 \frac{c\Delta t}{R}=y\left(1+\frac{uy}{8}
 +\frac{uy^2}{24}-\frac{u^2y^2}{112}\right).
 \label{eq:timedelay}
\end{equation}
The emission phase is shifted by $\Omega\Delta t$ before binning photons by
arrival phase.  The same delay convention must be used for every model in a
comparison; otherwise a purely procedural phase mismatch can dominate the
reported residual.  Circular spots are integrated as finite surface caps
rather than represented by their centers.  Two spots may have independent
colatitudes, longitudes, angular radii, temperatures, and beaming laws.

\subsection{Three prescriptions for the eccentricity}\label{subsec:shape}

We use three ways to set the surface flattening and compare the resulting
pulse profiles. A freely specified $e_\star$ shows how the profiles depend
on deformation, but is not, by itself, an equation-of-state prediction.

For a uniformly rotating, incompressible Newtonian fluid, the Maclaurin
relation is
\begin{equation}
      \varepsilon_{\rm rot}\equiv\frac{\Omega^2\Req^3}{GM}
      =\frac{3}{2e_\star^3}
      \left[(3-2e_\star^2)\sin^{-1}e_\star
            -3e_\star\sqrt{1-e_\star^2}\right].
      \label{eq:maclaurin}
\end{equation}
We invert Equation~(\ref{eq:maclaurin}) on the oblate branch. The Maclaurin
spheroid is an exact equilibrium shape for this idealized fluid, but its
uniform-density Newtonian assumptions can overpredict the flattening of
a realistic neutron star.

For the neutron-star surface, we use the fit of \citet{Morsink2007},
\begin{align}
      \frac{R_{\rm M}(\theta)}{\Req}
          & =1+a_0+a_2P_2(\cos\theta)+a_4P_4(\cos\theta),          \\
      a_0 & =-0.18\varepsilon_{\rm rot}
      +0.23\zeta\varepsilon_{\rm rot}-0.05\varepsilon_{\rm rot}^2, \\
      a_2 & =-0.39\varepsilon_{\rm rot}
      +0.29\zeta\varepsilon_{\rm rot}+0.13\varepsilon_{\rm rot}^2, \\
      a_4 & =+0.04\varepsilon_{\rm rot}
      -0.15\zeta\varepsilon_{\rm rot}+0.07\varepsilon_{\rm rot}^2,
      \label{eq:morsinkshape}
\end{align}
where $\zeta=GM/(\Req c^2)$.  We determine $e_\star$ by minimizing the
area-weighted squared radius difference between Equations~(\ref{eq:spheroid})
and (\ref{eq:morsinkshape}).  For the representative $M=1.43\Msun$,
$\Req=12\,$km, 300 Hz model, this procedure gives $e_\star=0.171015$,
with an area-weighted radius RMS difference of $6.03\times10^{-4}$ and a
maximum fractional difference of $1.04\times10^{-3}$.

\subsection{Hartle--Thorne spacetime and parameter relations}
\label{subsec:ht2}

Hartle--Thorne (HT) ray tracing provides the rotating-spacetime reference.
Its parameters are determined from the same stellar inputs used for the surface.
The exterior metric
retains all terms of order $O(J)$, $O(J^2)$, and $O(Q)$
\citep{HartleThorne1968}.  We define
\begin{equation}
 \begin{aligned}
 j&\equiv\frac{cJ}{GM^2},&
 \overline{Q}&\equiv-\frac{M Q_{\rm GH}}{J^2},\\
 q_{\rm HT}&\equiv-\frac{c^4Q_{\rm GH}}{G^2M^3}
 =\overline{Q}j^2.&&
 \end{aligned}
 \label{eq:htparams}
\end{equation}
where $Q_{\rm GH}<0$ for an oblate star.  In this convention the Kerr value
is $q_{\rm HT}=j^2$.  The ray calculation includes the linear frame-dragging
term, the spin-squared diagonal terms, and an independent quadrupole term.
The stellar surface is the same spheroid used in the OS calculations;
the HT coordinate radius at the equator is obtained from
$g_{\phi\phi}(r_{\rm eq},\pi/2)=\Req^2$.

For the neutron-star calculation, we obtain the dimensionless moment of
inertia from the compactness $C=GM/(\Req c^2)$ using the fit of
\citet{BreuRezzolla2016},
\begin{equation}
 \overline{I}=\frac{0.8134}{C}+\frac{0.2101}{C^2}
 +\frac{3.175\times10^{-3}}{C^3}
 -\frac{2.717\times10^{-4}}{C^4}.
 \label{eq:br16}
\end{equation}
The angular momentum is then
$j=\overline{I}(GM/c^3)\Omega$.  We infer the spin-normalized quadrupole
with the inverse I--Q fit of \citet{YagiYunes2013},
\begin{align}
 \ln\overline{Q}={}&-2.014+0.601\ln\overline{I}
 +1.10(\ln\overline{I})^2 \nonumber\\
 &-0.412(\ln\overline{I})^3
 +0.0459(\ln\overline{I})^4.
 \label{eq:yy13}
\end{align}
The Breu--Rezzolla and Yagi--Yunes relations and the Morsink surface fit
use the same $M$, $\Req$, and spin, but do not describe one
equation-of-state equilibrium model. We also use Maclaurin surfaces and
independently chosen $(e_\star,j,q_{\rm HT})$ values to examine the effects
of surface shape and spacetime parameters separately.

\section{Results} \label{sec:results}

We compare pulse profiles from spherical and oblate surfaces in a geometry
where their differences are visible. Three choices of eccentricity are
examined, with HT-ray used as the reference in each case.
\begin{figure*}
      \centering
      \includegraphics[width=0.96\textwidth]
      {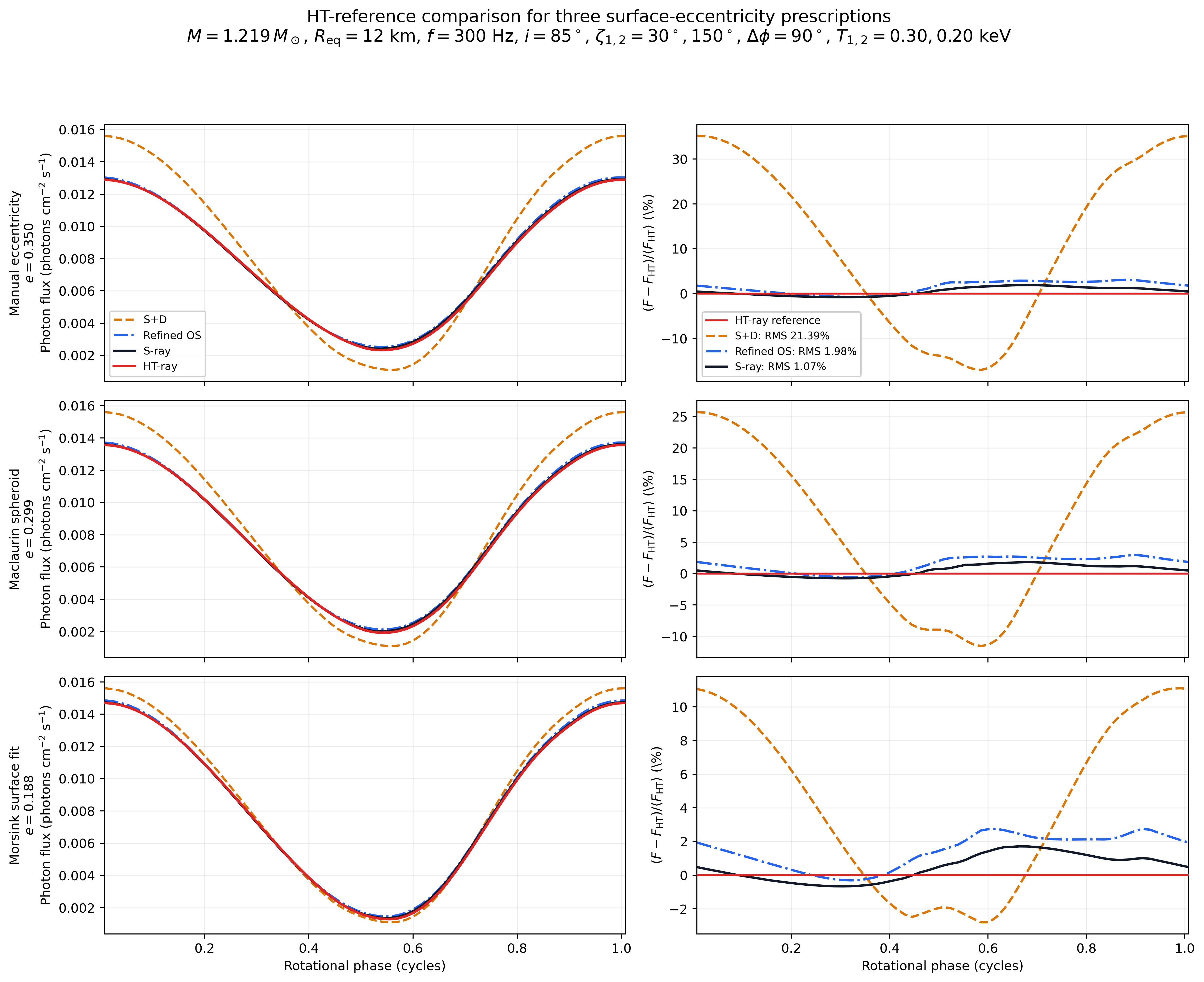}
      \caption{Pulse profiles (left) and residuals relative to HT-ray (right) for
            the manual, Maclaurin, and Morsink eccentricity prescriptions. The curves
            show S+D, refined OS, S-ray, and HT-ray.}
      \label{fig:eccentricitycomparison}
\end{figure*}

\begin{deluxetable*}{lrrrrrrr}
      \tablecaption{Profile Differences Relative to Hartle--Thorne Ray Tracing
            \label{tab:prescriptions}}
      \tablehead{
            \colhead{Surface prescription} & \colhead{$e_\star$} &
            \colhead{S+D RMS} & \colhead{S+D max} &
            \colhead{OS RMS} & \colhead{OS max} &
            \colhead{S-ray RMS} & \colhead{S-ray max}}
      \startdata
      Manual      & 0.35000 & 21.388 & 35.153 & 1.975 & 3.098 & 1.072 & 1.887 \\
      Maclaurin   & 0.29900 & 15.525 & 25.710 & 1.899 & 2.974 & 1.018 & 1.829 \\
      Morsink fit & 0.18796 &  6.503 & 11.105 & 1.756 & 2.748 & 0.917 & 1.711
      \enddata
      \tablecomments{The RMS and max columns are percentages of the
            phase-averaged HT-ray flux; max is the largest absolute phase residual.
            The manual and Maclaurin choices give more strongly flattened
            surfaces in these cases; the Morsink fit describes the
            neutron-star surface used for the physical comparison.}
\end{deluxetable*}

\begin{figure}
      \centering
      \includegraphics[width=\columnwidth]
      {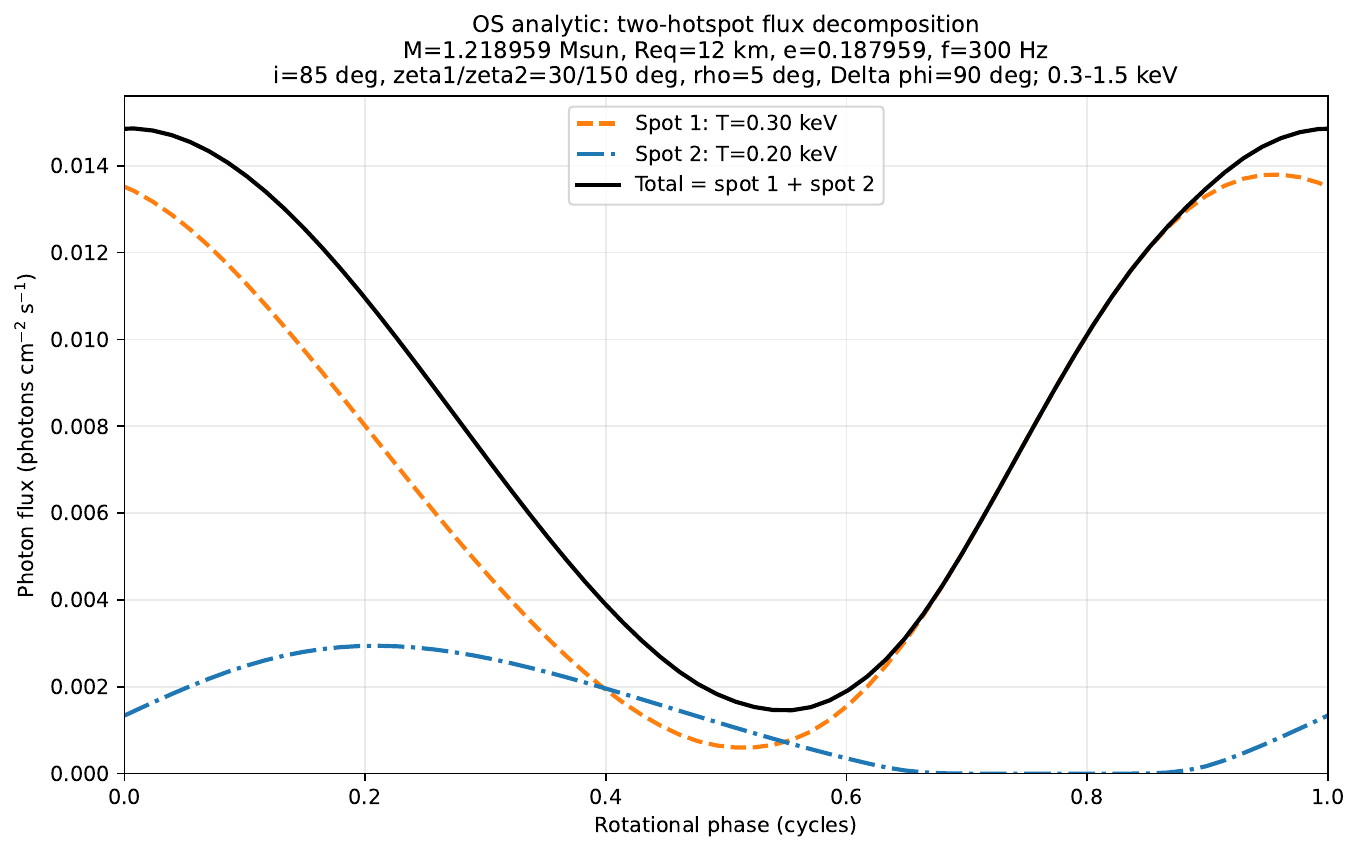}
      \caption{Refined-OS photon flux separated into the two hot-spot
            contributions for the Morsink surface. The total (black) is their
            sum. The cooler second spot contributes near the profile minimum
            but does not create a separate peak.}
      \label{fig:osdecomposition}
\end{figure}

\subsection{Selected geometry and numerical comparison}
\label{subsec:contrast}

A two-spot geometry with a clear oblate-surface signature exposes geometric
differences while the other stellar and spot inputs remain fixed.
The star has $2GM/(\Req c^2)=0.30$, $M=1.21896\Msun$, $\Req=12\,$km, and
spin frequency 300 Hz.  The observer inclination is $85^\circ$.  Two spots
of angular radius $5^\circ$ lie at colatitudes $30^\circ$ and $150^\circ$,
with a $90^\circ$ longitude offset and temperatures 0.30 and 0.20 keV.  The
flux is integrated over 0.3--1.5 keV.

The profiles are single-peaked despite the two spots: in the refined-OS
decomposition for the Morsink surface (Figure~\ref{fig:osdecomposition}),
the cooler second spot supplies only 15.6\% of the phase-averaged flux.
It affects the trough but is too weak to form a separate maximum.

For each surface prescription we compare S+D, refined OS, S-ray, and HT-ray. The last three models use the same spheroidal surface and spot geometry.
Schwarzschild ray tracing (S-ray) tests the analytic photon transfer;
second-order Hartle--Thorne ray tracing (HT-ray) is the reference for all
reported differences.  The relations for the spacetime parameters give
$j=0.171625$ and
$q_{\rm HT}=0.213172$.

For a profile $F_j$ sampled in $N$ phase bins, the tabulated RMS difference is
\begin{equation}
      \epsilon_{\rm flux}=\frac{1}{\langle F^{\rm HT}\rangle}
      \left[\frac{1}{N}\sum_{j=1}^{N}
            (F_j-F_j^{\rm HT})^2\right]^{1/2}.
      \label{eq:rmsflux}
\end{equation}
The figure shows $(F_j-F_j^{\rm HT})/\langle F^{\rm HT}\rangle$ at each
phase, and the table also gives its maximum absolute value.  No amplitude,
phase, or shape refitting is applied.  The ray profiles use 64 phase bins,
shared adaptive image-plane windows, nested grids, and $8\times8$ subpixel
coverage at spot boundaries; Appendix~\ref{app:convergence} gives the
resolution checks.

\subsection{Model comparison}\label{subsec:modelcomparison}

Comparing S+D, refined OS, and S-ray with HT-ray separates the effects of
surface shape, analytic photon propagation, and the rotating spacetime.
In the 200 Hz, single-spot example of \citet{Bogdanov2019}, the S+D and OS
profiles nearly overlap in the profile panel, although the spherical model
still has a measurable residual. In our selected 300 Hz, two-spot configurations,
the spherical S+D curve instead separates clearly from refined OS and HT-ray,
strongest for the
manually specified eccentricity and weakest for the Morsink prescription
(Figure~\ref{fig:eccentricitycomparison}). The refined OS curve follows the
ray-traced profiles much more closely in all three cases.
For the deliberately flattened $e_\star=0.35$ surface, the direct
S+D--refined-OS RMS difference reaches 20.7\% of the phase-averaged
HT-ray flux. Thus, the near-overlap in the 200 Hz example does not set
a general upper limit on the effect of oblateness.

The S+D RMS difference grows from 6.50\% for the Morsink surface to 21.39\%
for the most oblate manual surface (Table~\ref{tab:prescriptions}), showing
that the spherical approximation becomes less reliable as the deformation
increases.  Refined OS remains within 2\% RMS of HT-ray, while the
S-ray--HT-ray difference remains below 1.1\% RMS.
The latter represents the rotating-spacetime correction; the remaining
refined-OS--S-ray difference arises from the analytic photon-transfer
approximation.

\section{Conclusions} \label{sec:conclusions}
The near-overlap of S+D and OS profiles in the 200 Hz single-spot example of
\citet{Bogdanov2019} is not generic. For the selected 300 Hz configurations,
neglecting the latitude-dependent radius, proper surface element, oblate surface
normal, and spot visibility produces appreciable phase-dependent discrepancies.
Including these effects substantially improves agreement with ray tracing.
By contrast, the smaller difference between Schwarzschild
and Hartle--Thorne ray tracing indicates that frame dragging and the exterior
quadrupole are secondary corrections in the present cases.

The main contribution of this work is a refined analytic OS model that
captures these leading geometric effects while retaining the speed and
physical transparency of an analytic calculation. It provides an
intermediate description between the spherical S+D approximation and full
ray tracing, and is therefore suitable for rapid pulse-profile calculations
and future fitting applications. Extensions to realistic atmosphere emission
and numerical rotating-star spacetimes will be needed for more complete
parameter inference.

%% Please use the acknowledgment and contribution environments. This will 
%% be anonomyized when the "anonymous" style option is used. 



\FloatBarrier
\appendix
\section{Initially Ingoing Rays}\label{app:ingoing}

Initially ingoing rays require a separate analytic treatment, which
retains a geometric limitation.
The oblate normal can direct an outward-emitted photon initially toward
smaller Schwarzschild radius.  This branch is absent from the simplest
spherical outward-ray treatment.  Our analytic calculation uses the
first-order small-ingoing-angle correction of \citet{Morsink2007} for the
bending and propagation delay, matched continuously to the outgoing
analytic branch.  The current implementation applies this local
branch only when the surface-normal visibility test admits the photon.  It
does not yet trace the full path back against the oblate surface; a global
re-intersection rejection test is therefore retained as a limitation.

\section{Numerical Convergence and Reproducibility}\label{app:convergence}

Resolution checks on nested grids assess whether image-plane discretization
affects the reported residuals.
All calculations use 64 phase bins, the same adaptive image-plane window for
S-ray and HT-ray, exactly nested nodes, and $8\times8$ subpixel
coverage at hot-spot boundaries.  For the manual and Maclaurin prescriptions,
the $193^2$--$385^2$ checks give maximum HT-ray profile changes of 0.481\%
and 0.470\%, and maximum changes of the HT-referenced residuals of 0.492 and
0.476 percentage points.  For the Morsink prescription, the
$385^2$--$769^2$ check gives 0.162\% and 0.275 percentage points,
respectively.  Changes in the reported maximum residual remain below 0.12
percentage point in every case.

\section{Hartle--Thorne Internal Validation}\label{app:ht}

Independent tests of the HT metric and its derivatives complement the
profile comparisons.
Metric tests give maximum inverse-identity and analytic versus
finite-difference Christoffel errors of $6.94\times10^{-18}$ and
$8.52\times10^{-11}$.  The Schwarzschild, linear-in-$J$, spin-squared,
independent-quadrupole, and asymptotic component tests all pass;
Appendix~\ref{app:convergence} supplies the end-to-end resolution test.  The
remaining external check is to compare the HT rays and profiles with those
from a numerical rotating-star spacetime at the same equation of state and
stellar parameters.

\bibliography{MS}{}
\bibliographystyle{aasjournalv7}



\end{document}